\documentstyle[12pt]{article}
\begin{document}
\begin{center}
{\Large {\bf Exact metric around a wiggly cosmic string}} \\[15mm]
N. \"Ozdemir \\ [3mm]
ITU, Faculty of Sciences and Letters, Department of Physics,
80626 Maslak,  and 
Feza G\"{u}rsey Institute, 81220 \c{C}engelk\"oy;
Istanbul, Turkey \\[0pt]
\end{center}

\noindent

\begin{center}
{\bf Abstract}
\end{center}

\noindent
The exact metric around a wiggly cosmic string is found
by modifying the energy momentum-tensor of a straight infinitely
thin cosmic string to include an electric current along
the symmetry axis.

\newpage

Cosmic strings have widely been studied in the literature
over the last years \cite{vs}. There are two main reasons
for active investigations in the theory of cosmic strings;
the first is stimulated by cosmological implications
of the cosmic strings, as the string network in the early Universe
may provide viable density fluctuations for the formation
of the large-scale structure \cite{zel}. The second reason        
is due to unusual gravitational effects of cosmic strings \cite{vs}-
\cite{aho}, since the space-time geometry around
a straight cosmic string is locally flat, but globally it is
identical to the geometry of a cone \cite{his}. Numerical
simulations of the string network have shown the presence
of small-scale perturbations (wiggles) running along a long straight
cosmic strings \cite{vac}. These wiggles should affect on the
energy density and tension of the strings and can be taken into account
in the study of cosmological and physical consequences of the
cosmic string network. One way how to take into consideration
the effect of wiggles is to average over small-scale perturbations on
a long string. It leads to an effective energy density $\,\mu \,$ and
an effective tension $\, T\,$, which  satisfy to the equation of state
\begin{equation}
\mu T=\mu_0^2
\label{eqst}
\end{equation}
where $\,\mu_0 \,$ is the energy density per unit length of an unperturbed
cosmic string. With the equation of state (\ref{eqst}) the energy-momentum
tensor of a wiggly cosmic string oriented along the $z$-axis
can be approximated by the equation 
\begin{equation}
T_{\mu }^{\nu }=diag(\mu,0,0,T) \,\delta (x)\delta (y).
\label{em}
\end{equation}
Using this energy-momentum tensor as a source in the right-hand-side
of the linearized Einstein field equations one can find the gravitational
field around a wiggly cosmic string \cite{vil}. This result is given
by
\begin{eqnarray}
h_{00}=h_{33}=4G(\mu- T)\,\ln r \nonumber \\
h_{11}=h_{22}=4G(\mu+ T)\,\ln r
\label{gf}
\end{eqnarray}
where we can easily see that the wiggly cosmic string produces a newtonian
force. However it becomes not possible to extend the solution (\ref{gf})
to full relativistic case, that is to find the exact solution of
the Einstein field equations with the energy-momentum tensor (\ref{em}).

In this note we shall find the exact metric around a wiggly cosmic string
by modifying the energy momentum-tensor (\ref{em}). We shall suppose
that the wiggly string is still infinitely thin to be approximated
by the $\delta$-function like source, however unlike (\ref{em}), its
energy-momentum tensor contains additional components in the $x$
and $y$ directions. Thus we shall consider the energy-momentum tensor
\begin{equation}
T_{\mu }^{\nu }=diag(\mu +a,b,b,T + a)\,\delta (x)\delta (y)
\label{denerji}
\end{equation}
where $ a $ and  $ b $ are arbitrary constant parameters, which can
be related to the density of an electric current 
carried by the string along the $z$-axis.  We start with the metric
in cylindrical coordinates 
\begin{equation}
ds^2 = e^{2A} dt^2 - e^{2B} dz^2 - e^{2C} dr^2 -r^2 e^{2L}d\varphi^2
\label{metric}
\end{equation}
where $A, B, C$ and $L$ are the functions of distance $r$. Then the
Einstein field equations with the source (\ref{denerji}) are reduced
to the following set of equations
\begin{eqnarray}
(rB_{,r}{e^{A+B-C+L}})_{,r}=-4\pi ({\mu}-T+2 b)
{\delta(r)\over r}\\
((rL_{,r}+1){e^{A+B-C+L}})_{,r}=-4\pi (\mu+T+2 a)
{\delta(r)\over r}\\
{e^{A+B-C+L}}(rL_{,rr}+2L_{,r}+rL_{,r}^{2}-
C_{,r}(1+rA_{,r}+rB_{,r}+rL_{,r})       \nonumber \\
+\, rA_{,r}^{2}+rB_{,r}^{2}+rA_{,rr}+rB_{,rr})
=-4\pi({\mu}+{T}+2a){\delta(r)\over r}
\end{eqnarray}
where the subscript $_{,}r$ denotes derivative with respect
to $r$. The general solution which are subject to these equations
has the form
\begin{eqnarray}
A&=&2 (\mu-T-2 b)\ln r\\
B&=&-2 (\mu +T+2 b)\ln r\\
L&=&-\lbrack 2 (\mu+T+2a)+1\rbrack \ln r\\
C&=&-\lbrack 2 (\mu+T+4b+2a)+1\rbrack \ln r
\end{eqnarray}
where additive constants of integration are taken to be zero.
Finally, the metric is given by the expression
\begin{equation}
ds^{2}=\textstyle{(r^{4 (\mu-T-2 b)}dt^{2}}+
\textstyle{r^{-4 (\mu-T+2 b)}dz^{2})}-
\textstyle{r^{-4 (\mu+T+4 b+2 a)-2}(dr^{2}
+r^{2+16 b}d\varphi ^{2})}
\label{final}
\end{equation}
which describes the gravitational field near a wiggly cosmic string, carrying
a current on the $z$-axis. Indeed, the conservation of the energy momentum
tensor gives the relation
\begin{equation}
8 b^{2}+b(1-8 a-4 b(\mu+T))+2 ({\mu}-T)^{2}=0
\end{equation}
which enables us to reduce two unkwown constants $a$ and $b$ in the metric
to one of them, which in turn, as it was mentioned above,
can be related to the density of an electric current along the
symmetry axis \cite{mp}. We note that when the electric current
vanishes $ (a \rightarrow 0, b \rightarrow 0 ) $ the metric
(\ref{final}) reduces to its linearized limit \cite{vil}.

We have found the exact solution of the Einstein field equations
which determines the gravitational field of a long wiggly cosmic string
carrying an electric current on it. This is a generalization of a
corresponding linearized solution to the full relativistic case. It
requires the presence of a current on the symmetry axis of
the wiggly cosmic string.\\[5mm]
{\bf Acknowledgement}\\
I thank A. N. Aliev  for valuable discussions.

\end{document}